\documentclass[preprint,showpacs,preprintnumbers,amsmath,amssymb]{revtex4}
\usepackage{graphicx, latexsym, amssymb, amsmath, color, multirow, mathrsfs, booktabs, ifpdf}
\usepackage{dcolumn}
\usepackage{bm}

\def\phe6{$^6$He+$p$\ }
\def\he6pn{$p(^6$He,$^6$Li$^*)n$\ }

\newcommand{\beq}{\begin{equation}}
\newcommand{\eeq}{\end{equation}}
\newcommand{\beqn}{\begin{eqnarray}}
\newcommand{\eeqn}{\end{eqnarray}}
\newcommand{\bea}{\begin{array}}
\newcommand{\eea}{\end{array}}
\newcommand{\bsub}{\begin{subequations}}
\newcommand{\esub}{\end{subequations}}
\newcommand{\bpm}{\begin{pmatrix}}
\newcommand{\epm}{\end{pmatrix}}

\newcommand{\cals}[1]{{\mathcal #1}}

\newcommand{\svec}[1]{{\mbox{\boldmath${ #1}$}}}

\begin{document}
\title{Dirac-Brueckner Hartree-Fock Approach: from Infinite Matter to Effective Lagrangians for Finite Systems}
\author{Nguyen Van Giai$^1$}
\author{Brett V. Carlson$^2$}
\author{Zhongyu Ma$^3$}
\author{Hermann Wolter$^4$}
\affiliation{$^1$ Institut de Physique Nucl\'eaire, IN2P3-CNRS/Universit\'e Paris-Sud,\\
 91406 Orsay, France. \\
$^2$  Departamento de F\'isica, Instituto Technol\'ogico de Aeron\'autica, 12228-900 S\~ao Jos\'e dos Campos, Brazil \\
$^3$  China Institute of Atomic Energy, Beijing 102413, China \\
$^4$  Fakult\"at f\"ur Physik, Ludwig-Maximilians-Universit\"at M\"unchen, 
85748 Garching, Germany 
 }

\date{\today}

\begin{abstract}

One of the open problems in nuclear structure is how to predict properties of finite nuclei from the knowledge of a bare nucleon-nucleon interaction of the meson-exchange type. We point out that 
a promising starting point consists in Dirac-Brueckner-Hartree-Fock (DBHF) calculations using realistic nucleon-nucleon interactions like the Bonn potentials, which are able to reproduce satisfactorily the properties of symmetric nuclear matter without the need for 3-body forces, as is necessary in non-relativistic BHF calculations. However, the DBHF formalism is still too complicated to be used directly for finite nuclei. We argue that a possible route is to define effective Lagrangians with density-dependent nucleon-meson coupling vertices, which can be used in the Relativistic Hartree (or Relativistic Mean Field (RMF)) or preferrably in the Relativistic Hartree-Fock (RHF) approach. The density-dependence is matched to the nuclear matter DBHF results. 
We review the present status of nuclear matter DBHF calculations and discuss the various schemes to construct the self-energy, which lead to differences in the predictions. We also discuss how effective Lagrangians have been constructed and are used in actual calculations. We point out that completely consistent calculations in this scheme still have to be performed.

\end{abstract}
 \maketitle

\section{Introduction}
The Brueckner theory of nuclear many-body systems was
introduced about 50 years ago\cite{Brueckner54} as a powerful tool to relate
the bare nucleon-nucleon (N-N) interaction  V to the effective in-medium
interaction. Here and in the rest of this paper, we understand by bare interaction a meson-exchange type of interaction 
such as the Bonn potentials\cite{Machleidt89} which are designed for describing two-body data up to a few hundred MeV. 
This effective interaction takes into account the effects
of rescattering inside the nuclear medium and it is generally known as
the Brueckner G-matrix. The G-matrix differs from the scattering
T-matrix of an isolated N-N pair because of the presence of the
surrounding nucleons which limits the available phase space. Thus, the
 G-matrix deviates from the bare
interaction V, in particular by the fact that G behaves smoothly
whereas V is strongly repulsive at short distances. Then, it becomes
conceivable that a nuclear system can be described by an ensemble of
nucleons moving in a mean field created self-consistently by the
two-body effective interaction G while this picture would not hold in
a simple Hartree-Fock (HF) description based directly on V. The
Brueckner-Hartree-Fock (BHF) theory\cite{Brueckner55,Bethe56} is thus an essential
link between the bare N-N interaction V and the properties of nuclear
matter and atomic nuclei.

However, it was found already in early BHF calculations \cite{Erkelenz74} that the equation of state (EOS) 
calculated in BHF theory does not lead to the empirical saturation point
of symmetric nuclear matter ( $\frac{E}{A}\simeq$-16 MeV at
$k_F\simeq$1.36 fm$^{-1}$). Indeed, it is well-known that all BHF
predictions of the nuclear matter saturation point lie along the
so-called Coester band \cite{Coester70} which misses the empirical
saturation point. It is only when contributions of
3-body forces are taken into account that the saturation point can be
satisfactorily predicted \cite{Baldo99,Zuo02}. A large part of these 3-body terms were later understood as effects of the vacuum, i.e., virtual nucleon-antinucleon excitations, leading to the so-called Z-graphs in the diagrammatic expansion \cite{Brown87,Bouyssy87}. 

It is then reasonable to include effects of relativity from the beginning. 
The relativistic version of the BHF theory is known as the
Dirac-Brueckner-Hartree-Fock (DBHF) approach since the early work of 
Refs.\cite{Anastasio83,Brockmann84} and the fully self-consistent
nuclear matter calculations of Ter Haar and 
Malfliet \cite{TerHaar87}. The starting point of the DBHF theory is a
realistic bare N-N interaction formulated in terms of meson exchanges, the parameters of which are adjusted to reproduce N-N scattering data in the T-matrix approximation. This One-Boson-Exchange (OBE) interaction is used in a relativistically covariant framework
by solving the Bethe-Salpeter (BS) equation, or more precisely a
Thompson equation obtained by a covariant three-dimensional reduction
of the BS equation \cite{TerHaar87}. 
In the relativistic approach the nuclear mean field is obtained by a
large cancellation between scalar and vector potentials.
The scalar potential saturates at high
densities, thus providing an additional saturation mechanism. 
It is then found that the DBHF approach based on
meson-exchange potentials predicts a nuclear matter saturation point
away from the Coester band without the need for a 3-body force
and in relatively good agreement with the
empirical point \cite{TerHaar87,Brockmann90}. 
This advantage of DBHF over
the non-relativistic BHF makes the applications to finite nuclei
practicable in principle and, in fact, more simple.

The Brueckner theory is technically difficult, which is related to the double self-consistency
of the problem: the determination of the effective in-medium interaction G requires the knowledge of the
self-consistent single-particle mean field, but this mean field
itself depends on G. This double self-consistency can be handled in
simple cases like homogeneous infinite nuclear matter but it
becomes very cumbersome and difficult to treat in finite nuclei. There are some BHF calculations of finite nuclei, e.g., Ref.\cite{Anastasio78},  but not including 3-body forces, and only very rare calculations in the relativistic DBHF  version\cite{Muether90}.

Then the question arises whether one can effectively perform self-consistent mean field calculations for finite nuclei  based on the knowledge of the mean field properties of infinite nuclear matter deduced from BHF or DBHF calculations. 
The idea is to build an effective Hamiltonian that can reproduce the results of the non-relativistic BHF (plus corrections), or the relativistic DBHF  in infinite nuclear matter in simpler many-body schemes, namely the non-relativistic Hartree-Fock (HF) or the relativistic Dirac-Hartree-Fock (DHF) approximations, 
and then solve the HF or DHF equations for finite nuclei.
This issue was first taken up in the 1960's and the
work of Negele\cite{Negele70} showed that it was indeed possible to
describe finite nuclei reasonably well using a G-matrix coming from
infinite matter calculations, at the cost of some readjustments which were attributed to the neglect of 3-body effects and higher order terms.  

In the light of the success of nuclear matter DBHF calculations it then seems most promising to look for 
effective Lagrangians which are written in terms of OBE interactions with density-dependent meson-nucleon coupling vertices. The nuclear matter DBHF results would have to be reproduced in DHF approximation by these Lagrangians. Then, these adjusted Lagrangians would be used for describing finite nuclei in the DHF approximation. So far, this program has never been carried out, and this constitutes the open question which is addressed in this paper.  

This scheme is in the spirit of a density functional approach, since it leads to an energy functional in terms of the baryonic density alone, which contains the effect of exchange and correlations. The application of such a Lagrangian to finite nuclei implies a sort of Local Density Approximation (LDA). This has been justified in noticing that the range of the effective interactions due to meson exchanges are smaller that the scale of spatial variations in a nucleus\cite{Negele70}. This assumption is weakest for $\pi$-meson exchange, but this field has special properties anyway, because of its pseudo-scalar (PS) nature. It should also be understood, that these meson exchanges do not necessarily represent the exchange of physical in-medium mesons, but they rather represent the different Lorentz components of the decomposition of the DBHF self-energy, such as  scalar/vector or isoscalar/isovector fields. One should therefore rather speak of meson-like fields. In this spirit the masses of these various mesons do not necessarily have to be the masses of the physical, even less the free, mesons, 
even though this has been the usual assumption.

The density dependence of the effective Lagrangian (i.e. of the coupling vertices) is determined by comparison of the relativistic Hartree (also called the Relativistic Mean Field (RMF)) or the relativistic Hartree-Fock (also called Dirac HF (DHF)) approaches with the full DBHF results. 
The question is then to what physical quantities of DBHF they are matched. This question arises because in practice the DBHF equations are often solved in a three-dimensional Thompson reduction, also neglecting the negative energy states (no-sea approximation). Then, the Lorentz structure of the 
self-energy is either lost or ambiguous. This will be discussed in the next section.

The infinite matter DBHF calculations, on which this scheme is based, have limitations at very low and very high densities. At very low densities they become instable because of the deuteron pole, or more generally because of cluster effects and non-inhomogeneities of nuclear matter \cite{Typel09}. At densities several times the value $\rho_0$ of saturation density the description in terms of nucleons and mesons may not be valid any more, and more nucleon and meson states (also strangeness degrees of freedom), and eventually subnuclear degrees of freedom have to be included. This is the regime of more or less energetic heavy ion collisions, where these questions are also of immediate relevance. They are also of relevance in the structure of neutron stars and core-collapse supernovae, where the question of the density dependence of the symmetry energy enters in particular. However, in the regime of nuclear structure, i.e., in the range of about $0.1\rho_0$ to $(1-2)\rho_0$ the approach sketched above may be valid. It is this approach which we want to review in this article.

The structure of this article is the following. In Sect. II we shall review the status of DBHF calculations for infinite nuclear matter, in particular the problem of constructing the various Lorentz components of the self-energy, also discussing asymmetric nuclear matter. In Sect. III we discuss the question of constructing effective (density-dependent) Lagrangians from the DBHF solutions in nuclear matter.
We also make sime comments on applications of such effective Lagrangians to finite nucleus calculations. 
Concluding remarks are made in Sect. IV. We hope it will be apparent, that on one hand the scheme of effective DBHF Lagrangians is an important and viable route from the realistic N-N interactions to realistic calculations of finite nuclei. Compared to the highly successful phenomenological RMF and RHF calculations this approach 
is essentially parameter-free, in the sense that there are no parameters adjusted to nuclear matter or finite nuclei.
On the other hand, we will see that a completely satisfying solution to this problem is still to be found.

\section{Status of DBHF calculations in infinite matter  }

The earliest study of infinite nuclear matter based on the self-consistent
framework of DBHF theory
was performed by Erkelenz\cite{Erkelenz74}. As his work also explored
the description of the N-N scattering data using a one-boson exchange
(OBE) interaction, he compared the results obtained using several
early versions of the OBE without attempting to reproduce the expected
saturation properties of nuclear matter. With the success of Dirac
phenomenology in the description of proton-nucleus scattering, the
problem was taken up again in Ref. \cite{Anastasio83}, where reasonable
agreement with the expected nuclear matter properties were obtained
using the OBE N-N interactions available at the time. In the pioneering
work of Ter Haar and Malfliet\cite{TerHaar87}, a one-boson exchange 
interaction reproducing N-N and N-$\Delta$ data in the energy
range up to 1 GeV was used in DBHF calculations of symmetric nuclear
matter and the effects of the N-$\Delta$ degree of freedom were also
studied. In the meantime DBHF calculations have reached a high level of sophistication. They confirm that, starting from high precision N-N interactions, one may obtain good predictions for nuclear matter. This also gives confidence that this approach can be used to extrapolate to extreme states of nuclear matter at high density and temperature or high asymmetry. Such predictions are important for investigations in exotic nuclei, heavy ion collisions, and astrophysics.

However, DBHF calculations do not yield identical results, even when the same N-N interactions are used.
This has its origin in the difficulty mentioned earlier of the requirement
of double self-consistency of both the G-matrix and the single-particle mean
field in solving the Bethe-Salpeter equation. To simplify
its solution, the reduced Bethe-Salpeter equation is usually projected
on the positive-energy Dirac spinors, which yields an equation very
similar to the non-relativistic one and permits a similar
solution. However, the Lorentz structure of the G-matrix is lost in such a
reduction and must be reconstructed in some manner. Various approximation schemes have been used to deal with this problem. This not only affects the results of the DBHF calculations itself, but also the construction of effective Lagrangians to be used for finite nuclei calculations, as discussed in the next section.

From parity and time reversal symmetry the most general form of the self-energy in nuclear matter has the form
\beq
\Sigma(\svec k) = \Sigma^s(\svec k) - \gamma^0 \Sigma^0(\svec k)+ 
\svec\gamma\cdot\hat{\svec k} ~ \Sigma^v(\svec k)  ~,
\label{se}
\eeq 
with the scalar, time and space-like vector components of the self-energy. In asymmetric nuclear matter these are different for protons and neutrons and the  appropriate isoscalar and isovector components have to be determined in addition. In the Hartree and also in the HF approximation these follow in a unique way from the exchange of the corresponding meson fields. In DBHF, however, the exchange and correlation diagrams mix the contribution of the various mesons to the components of the self-energy. Thus, $\pi$-meson exchange, for example, contributes to the self-energy even though there is no pseudo-scalar (PS) self-energy in Eq.(\ref{se}). The above remark also explains why in the Introduction we made the comment that one should talk of meson-like DBHF fields and not of the exchange of physical mesons. Thus, the Lorentz structure has to be reconstructed, giving the correct contribution of the various meson exchanges.

The simplest method to determine the above components of the self-energy was introduced by Brockmann and Machleidt \cite{Brockmann90,Bao94} and consists in analyzing the momentum dependence of the single-particle energies, 
\beq 
\epsilon(k)=\sqrt{k^2+M^{*2}} +\Sigma_0~, \label{relspe}
\eeq
where the effective mass is $M^*=M+\Sigma^s$. It was later found\cite{Shen97,Ulrych97} 
that, while this method gives
reasonable results for symmetric nuclear matter, it leads to incorrect
results in asymmetric matter, even giving the wrong sign for the isospin dependence of the effective interaction. The alternative approach is to project
the G-matrix elements on a complete set of five - in asymmetric matter
six - Lorentz invariants\cite{Horowitz87}. However, the choice of these invariants is
not unique. Different choices give identical matrix elements for the Hartree self-energies for on-shell positive energy states, but the exchange and correlation contributions, as well as the coupling to negative energy states, can be very different leading to different self-energies. Thus, various projection schemes have been developed which differ mainly in the effect of the PS meson exchange.

One possible approach is to use the PS operator as one of the invariants. The
pseudo-vector (PV) pion-nucleon coupling is defined so as to yield the same matrix
elements between positive energy states as the PS 
coupling. A PS coupling, however, furnishes very strong
coupling between positive and negative energy states and it is unsuitable
for use in nuclear matter calculations. As the covariant
reconstruction cannot distinguish between the PS and
PV forms, it can furnish unexpectedly large and undesired
PS contributions. 
It also leads to a very strong momentum dependence of the self-energies, see e.g., Ref.\cite{Dalen04}. 
Therefore, the preferred choice has been to use a pseudovector PV invariant. However, this is also not completely satisfactory, because the $\rho$-exchange for instance is not treated correctly then\cite{Dalen04}. A compromise, which has proven to be most successful, is a decomposition of the DBHF G-matix of the form
\beq G = V +\Delta G ~, \label{deltaG}  \eeq
where V is the starting OBE interaction. This approach was introduced by the group of M\"uther \cite{Ulrych97,Schiller01}, and in a slightly different form as the 'optimal T-matrix representation' by the group of Fuchs \cite{Dalen04,Dalen05,Dalen07}. Since the Lorentz structure can be derived explicitly in the HF approximation for the OBE interaction, the first part
of $V+\Delta G$ is treated exactly. Only the difference $\Delta G$ is treated by a projection method. This method seems to yield the most reliable results at present.
In this method $\Delta G$, which represents mainly the correlation effects, has been parametrized in terms of effective meson exchanges of small or zero ranges with density dependent coupling vertices. In a related approach Boersma et al.\cite{Boersma94a,Boersma94b} have parametrized the G-matrix in terms of Yukawa function with fitted masses and coupling strengths including a form factor. These effective interactions have then been used in finite nucleus calculations directly (see next section). 

The PV projection approaches with or without splitting off the bare OBE interaction usually agree rather well for symmetric nuclear matter around saturation density. They give differences, however, for asymmetric or neutron matter, and also for higher densities. Thus, these different choices are of importance for the structure of exotic nuclei with neutron-rich low-density skins and also for heavy ion collisions and the structure and mechanisms of astrophysical objects. 

Still another method consists in performing the projection on the full space of positive and negative energy Dirac states, as proposed e.g. by de Jong and Lenske \cite{Jong98a} and also in a slightly different framework by Huber at al.\cite{Huber95}. There would then be no ambiguity. However, the authors of Ref.\cite{Jong98a} point out that, in order  for this scheme to be satisfactory, the N-N T-matrix would have to be determined in the same scheme, which is not usually the case. Thus, this method also needs further development.

It is well known that the DBHF is not a thermodynamically consistent
approximation \cite{Baym61}. Several attempts have been made to go
beyond it by developing a Dirac T-matrix approximation to nuclear
matter\cite{Jong91, Jong96}. These yield results in agreement with
previous ones near saturation but tend to predict a stiffer equation of
state at higher densities. A calculation explicitly including the
negative energy states has also been performed \cite{Jong98a} and it also yields a
stiffer equation of state at high densities. However, these
calculations require approximations unnecessary for the DBHF, such as
the quasi-particle approximation, and should be improved before
conclusions are made about their high density behavior.

Other approximations and difficulties are relevant to the
DBHF. Schiller, M\"uther and Czerski\cite{Schiller99} have shown that binding energies
near saturation in Brueckner calculations can vary by about 10\% due to angle
averaging of the Pauli blocking factor. Clustering creates serious problems
for obtaining convergent solutions at low nuclear matter densities
and, even where convergent, furnishes binding energies that cannot be
described by a Hartree-Fock type calculation\cite{Margueron07}. At
sufficiently high densities, one can also question the validity of a
OBE approximation to the interaction, as discussed in the Introduction. 
Yet, in spite of the many open
questions, the existing nuclear matter DBHF calculations indicate that
one can describe the bulk properties of infinite matter in terms of 
realistic OBE interactions, at least at values of matter density 
relevant for nuclear structure calculations, which 
are neither too low nor too high.

\section{DBHF-based effective Lagrangians}

The next step is to determine effective Lagrangians which can be used in a self-consistent mean field approach like RMF or RHF and which reproduces the DBHF results in homogeneous matter. The effective Lagrangians is 
described as a system of nucleons of Mass $M$ and effective mesons interacting via meson-nucleon couplings which are determined by the coupling strengths $g_i$ in the following form
\beqn
\mathcal{L} & = & \bar{\psi}\left( i \gamma_{\mu}\partial^{\mu} - M \right) \psi_{i} + \mathcal{L}_{mes}^0(\sigma,\omega,\vec{\rho},
\vec{\delta},\pi,A) + \mathcal{L}_{int} \\
\mathcal{L}_{int} & = & g_{\sigma} \bar{\psi}\sigma\psi
- g_{\omega} \bar{\psi}\gamma_{\mu}\omega^{\mu}\psi
- g_{\rho} \bar{\psi}\gamma_{\mu}\vec{\rho}^{\mu}\cdot\vec{\tau}\psi
+ g_{\delta} \bar{\psi}\vec{\delta}\cdot\vec{\tau}\psi \\ \nonumber
& + & \frac{f_{\pi}}{m_{\pi}} \bar{\psi}\gamma_5\gamma_{\mu}\partial^{\mu}
\vec{\pi}\cdot\vec{\tau}\psi
- e \bar{\psi}\gamma_{\mu} \frac{1}{2}(1+\tau_3) A^{\mu}\psi \quad ,
\label{QHD}
\eeqn
with the fields of the nucleon ($\psi$), two isoscalar meson fields ($\sigma$ and $\omega$), three isovector fields ($\vec{\pi}$, $\vec{\rho}$ and $\vec{\delta}$) and the photon ($A$). 
The free Lagrangian $\mathcal{L}_{mes}^0(\sigma,\omega,\vec{\rho},
\vec{\delta},\pi,A)$ of the mesons and the photon has the usual form with masses $m_i$ of the mesons, which are often assumed to be the masses of the corresponding free mesons, even though this does not have to be the case considering the effectiveness of the meson fields. 
In many models the $\delta$-degree of freedom is omitted and the isovector properties rely only on the $\rho$ meson if the RMF (Hartree) framework is adopted. 
In this case the pion is also omitted since it cannot contribute to the Hartree energy due to parity conservation. 
On the other hand, in RHF all mesons contribute to isovector properties, even isoscalar mesons through their exchange contributions to self-energies.

The stationarity condition of the action integral $\int d^4x \cals L(x)$ under variations of the
physical fields $\phi$ ($\phi = \psi, \sigma, \omega, \vec{\rho}, \vec{\pi}, \vec{\delta}$, and $A$) leads to the Euler-Lagrange equations, from
which one can derive the equations of motion for the meson, photon and nucleon fields \cite{Bouyssy87}. The meson
and photon fields obey inhomogeneous Klein-Gordon equations and a Proca equation with source terms, respectively,
whereas the nucleon field obeys a Dirac equation coupled minimally to the meson fields. In homogeneous matter they have simple analytic forms which make the solving of the self-consistent problem much easier than in finite nuclei.
 It is customary to make the simplifying assumption of neglecting the time component of the four-momenta
carried by the mesons, i.e., the meson fields are assumed to be time independent. This amounts to neglect the
retardation effects. The nuclear energies involved are small compared to the masses of the exchanged mesons, so that this
approximation should be valid for the $\sigma$-, $\omega$-, $\delta$- and $\rho$-induced interactions, and also, to a lesser
extent, for the pion.

The meson-nucleon couplings must be determined by reproducing as closely as possible the DBHF results of infinite matter in some range of values of baryonic density and neutron-to-proton ratio. 
Since the DBHF self-energies are density and momentum dependent, the coupling vertices necessarily have to be density dependent, (we will discuss the momentum dependence below), 
i.e., one has $g_i(\rho_B)$ for the different meson species $i$, where $\rho_B$ is the baryon density. 
Of course, the various approximations schemes for the self-energies of DBHF in nuclear matter, which were discussed in sect. II, will lead to different results also for the effective Lagrangians. As discussed there, the method of extracting these self-energies from 
their momentum dependence
\cite{Brockmann90} was shown to give incorrect asymmetry properties\cite{Ulrych97}. Thus, we will concentrate on 
approaches which are based on matching the Lorentz components of the self-energy. Finally, there are also the approaches based on parametrizations of the DBHF G-matrix directly.

Two scenarios have been applied with respect to the approximation in which the DBHF self-energies are matched: either in the Hartree or the HF approximation of nuclear matter. In the former case this has been called the density-dependent relativistic Hartree (or mean field) approximation (DDRH or DDRMF), in the second the DDRHF approximation. 
In DDRH the result is simple: the density-dependent coupling vertices are given via the ratio of the DBHF self-energies divided by the respective Lorentz components of the density, e.g.,
\beq
\left ( \frac{g_{\sigma}}{m_{\sigma}}\right )^2 = \frac{1}{2}\frac{\Sigma^s_n(k_{Fp},k_{Fn})
+ \Sigma^s_p(k_{Fp},k_{Fn})}{\rho^s_n + \rho^s_p} \quad ,
\label{effcoupl}
\eeq
and correspondingly for the other coupling coefficients (those for the $\pi$-coupling are given in ref.\cite{Shi95}). In Eq.(\ref{effcoupl}) we have explicitly indicated that the coupling vertices depend on the neutron and proton densities separately (resp. the isoscalar and isovector densities). In practice it was often found, that the dependence on the isovector density is weak and this dependence has been neglected 
(see, 
e.g., ref.\cite{Hofmann01}). In the Hartree-Fock approximation (DDRHF) the expressions which have to be matched are slightly more complicated, and are given also in ref.\cite{Shi95}. 

The DBHF self-energies are explicitly momentum dependent due to the contributions of the exchange and correlation terms (the exchange contribution follows roughly a $1/k$ dependence). In the DDRH approach, however, the RH self-energies in nuclear matter are momentum independent, and in the fitting procedure some approximation has to be made. The approach followed, was either to match the self-energies at the Fermi surface (e.g.\cite{Brockmann92}) or to average over the Fermi sphere\cite{Shi95}. Another approach has been to perform a first order correction involving the derivative of the self-energies with respect to momentum at the Fermi surface \cite{Hofmann01}. A covariant way to parametrize the momentum dependence in phenomenological RMF models has been suggested by Typel\cite{Typel09}, but has not yet been used to parametrize DBHF self-energies.

In the DDRHF approach the RHF expressions in nuclear matter are density dependent 
(roughly as $1/k$) and thus the matching to the DBHF self-energies can be done in a more consistent way. In principle the DBHF self-energies also carry a momentum dependence from the correlation terms, but the matching to the $1/k$ dependence usually is quite accurate, see e.g. ref.\cite{Shi95}. Thus, the DDRHF approach should be preferred and should give a better connection between nuclear matter and finite nuclei. Fock exchange terms can be quite important in finite nuclei, as they can modify the shell structure or avoid unphysical shell gaps \cite{Long07}. RHF calculations in finite nuclei are more involved than RH ones but the additional technical difficulties can be solved\cite{Bouyssy87}. 

Of course, the parametrization of the coupling vertices and the approximation in the calculation of finite nuclei has to be consistent. 
A first check has been that the equation of state, i.e., the energy per particle as a function of density, of DBHF is reproduced with the 
density-dependent parametrization in the same approximation (DDRH or DDRHF).
A  parametrization of DBHF in the Hartree approximation should be used in DDRH (DDRMF) calculations of finite nuclei, and the one obtained in RHF should be used with DDRHF calculations. 

Finally the G-matrix of DBHF calculations has been parametrized in terms of the exchange of effective mesons by Schiller et al.\cite{Schiller01} or by Boersma and Malfliet \cite{Boersma94a,Boersma94b}. These effective interactions have then been used directly in a finite nucleus calculation \cite{Fritz93,Fritz94} or 
used to reconstruct a self-energy 
which is used in turn to build a density-dependent effective Lagrangian 
\cite{Ma02}. In this last procedure, however, the RHF method should be used for consistency. 

A Lagrangian as in Eq.(\ref{QHD}), in which the coupling strengths depend explicitly on the density $\rho_B$ violates the covariance of the model. 
This has been treated by writing the density dependence in terms of scalar invariants built from nucleon field operators related to the density\cite{Fuchs95}. The form of this operator is not a priori given, and different choices have been investigated, e.g., the scalar or the square of the vector density operator. The dependence on field operators necessarily leads to additional terms in the Euler-Lagrange equations of motion, so-called rearrangement terms, which depend on the type of density dependence assumed.  In the investigations of Fuchs et al.\cite{Fuchs95} it was found that a vector-density dependence gave better results for finite nuclei starting from the same effective DBHF self-energies, and such a dependence is now generally employed. 
In any case rearrangement contributions should always be included. It was shown in Ref.\cite{Fuchs95} that they are necessary for the thermodynamic consistency of the effective theory. The effect of the rearrangement terms in phenomenological density-dependent models has recently been studied in detail in Refs.\cite{Long06,Long08}.

Attempts have been made to calculate the properties of finite nuclei with the above mentioned effective Lagrangians.
These studies have been done in the DDRMF\cite{Brockmann92,Fuchs95,Shen97,Hofmann01,Ma02} or in the DDRHF approaches\cite{Boersma94b,Fritz94,Shi95}. Consistent rearrangement contributions were included in Refs.\cite{Fuchs95,Shen97,Hofmann01,Ma02}, but not in Refs.\cite{Brockmann92,Fritz94,Boersma94b,Shi95}. From the arguments presented in this article a complete calculation should be done in DDRHF including rerrangement terms and matching to DBHF self-energies, which are obtained from a "subtracted" G-Matrix like in Refs.\cite{Schiller01,Dalen07}. Looking through the calculations existing at present such a calculation does not yet exist, even though all the ingredients seem to be present. 

It is not the place here, to compare in detail the different results of the finite nucleus calculations based on DBHF calculations. Today, the focus is especially on the isovector properties and the extrapolation to very neutron-rich nuclei. Generally it can be said that the quality is comparable to the results of the phenomenological DDRMF or DDRHF calculations, where the density dependent parameters are adjusted directly to data of nuclear matter and finite nuclei with a number of (global) parameters ranging from 7 to 10, which naturally lead to a greater precision with respect to the data. Recently, the calculation scheme for a complete relativistic Hartree-Fock-Bogoliubov (RHFB) description of finite nuclei has been proposed\cite{Long09,Ebran09} and the results are at an excellent quantitative level using a phenomenological Lagrangian.

\section{Conclusion}
In this work we have addressed the question of how to link the bare N-N interaction V to the properties of infinite and finite nuclear systems. This issue is of utmost importance because of the need to have a theory capable of predicting nuclear properties in regions of the nuclear chart not yet accessible to experiment. The self-consistent mean field is a good approach, but its predictive power cannot be extrapolated very far if one relies only on phenomenological adjustments of effective interactions or energy density functionals, as it is done in non-relativistic approaches. There are also attempts in non-relativistic mean field studies to use the bare N-N interaction limited to some finite part of momentum space,  for instance with the $V_{\rm{low-k}}$ interactions. Here, we have discussed the possibility to deduce the effective N-N interaction via Brueckner theory. This could be done in a non-relativistic approach or a relativistic one. The main advantage of the relativistic Brueckner theory, or DBHF, is that the nuclear matter equation of state is satisfactorily obtained without adding explicitly the effects of 3-body forces as it is the case for the non-relativistic Brueckner theory. Thus, the DBHF equation of state and self-energies are good starting points for deducing effective Lagrangians to be used in relativistic self-consistent studies. 

With respect to the present state of the subject, there are not yet completely satisfactory RHF calculations of finite nuclei based on effective Lagrangians deduced from DBHF nuclear matter studies. The main difficulty was identified as coming from the ambiguity in the choice of the covariant operators on which the effective G-matrix has to be expanded. A way to get around has been proposed by Schiller and M\"uther\cite{Schiller01} who suggest to split $G$ into $V+\Delta G$, the idea being that $V$ is the dominant part of $G$ and thus, a large part of the ambiguity can be eliminated. Indeed,  there have some attempts to build $G$ in this way\cite{Ma02} but the use of such parametrized, density-dependent G-matrix in RHF calculations of finite nuclei is still to be done. 

With the recent progress in RHF and RHFB calculations using density-dependent meson-nucleon couplings, the next step to establish a link between N-N interactions like the Bonn potentials and the properties of finite nuclei is at hand. This is one of the open problems in nuclear structure which need to be solved, and the route through DBHF and density-dependent RHF seems a promising option.


\end{document}